\documentclass[10pt,a4paper,twoside]{article}

\usepackage{indentfirst}			
\usepackage{graphicx}				
\usepackage{amsgen,amsfonts,amssymb,amsbsy}	

\newenvironment{resum}{\begin{quote}\small}{\end{quote}}

\newcommand{\bfsf}[1]{\textsf{\textbf{#1}}}

\begin{document}

\thispagestyle{plain}		

\begin{center}


{\LARGE\bfsf{Towards an open set of regular cosmological models}}

\bigskip


\textbf{L.\ Fern\'andez-Jambrina}$^1$ and \textbf{L.M.\ Gonz\'alez-Romero}$^2$


$^1$\textsl{Universidad Polit\'ecnica de Madrid, Spain.} \\
$^2$\textsl{Universidad Complutense de Madrid, Spain.}

\end{center}

\medskip


\begin{resum}
The possibility of obtaining an open set of regular cosmological 
models is discussed. Cylindrical stiff perfect fluid cosmologies are 
studied in detail. The condition for geodesic completeness is easy to 
check. A large family of non-singular models is found 
therein.
\end{resum}

\bigskip


\section{Introduction}

Just to settle the issue from the beginning, by a regular 
cosmological model we consider a perfect fluid cosmological model 
which is causally geodesically complete, that is, every lighlike or 
timelike geodesic can be extended from minus infinity to infinity in 
the affine parametrization.

These models have been neglected in the past since the powerful 
singularity theorems by Hawking, Penrose and Tipler (see, for 
instance, \cite{HE, Beem}) seemed to point out that they would 
violate some physical requirements (causality conditions, energy 
conditions). For instance, it was thought that they would enclose 
closed timelike curves or negative energy densities. However, twelve years ago 
Senovilla published the first known regular 
model for a cylindrical universe with a radiation fluid as matter 
content \cite{Seno}.

These models did exist then, but they were considered an issue of 
luck. Further results  seemed to support these thoughts, since the 
list of regular models is rather sparse.

For instance, in the Ruiz-Senovilla family, regular models are a just 
a zero-measure set \cite{esc}.

Stiff perfect fluids are an excellent arena for checking conjectures 
of this sort, since Einstein equations can be integrated almost to 
the end.

\section{Inhomogeneous stiff fluid cosmologies}

We shall consider inhomogeneous spacetimes with cylindrical symmetry 
and, as a matter of convenience for integration, we choose 
ignorable coordinates for the Killing fields and isotropic coordinates 
for the rest. Metric functions depend just on $t$ and $r$ then.

\begin{equation}
ds^2=e^{2K}(-dt^2+dr^2)+e^{-2U}dz^2+\rho^2e^{2U}d\phi^2,\label{metric}
\end{equation}
\begin{equation}
    -\infty<t,z<\infty,\ 0<r<\infty,\ 0<\phi<2\pi.
\end{equation}

Einstein equations are written in a comoving frame for the fluid, 
$u\propto \partial_{t}$.

\begin{equation}
    T^{\mu\nu}=\mu u^{\mu}u^\nu+p\,(g^{\mu\nu}+u^\mu u^\nu),\qquad 
    0\le\mu,\nu\le 3,\qquad u^\mu u_{\mu}=-1,
\end{equation} 
where the energy density, $\mu$ and the pressure, $p$,  are equal for 
a stiff perfect fluid.

The only assumption we make, since every regular cosmological model 
in the literature has a spacelike transitivity surface element, 
$\rho$, is imposing that $\rho$ must be orthogonal to the velocity of 
the fluid $u$.

Rescaling the coordinates we may take $\rho=r$ and the system is 
quite simple,
\begin{eqnarray}
U_{tt}-U_{rr}-\frac{U_{r}}{r}=0,\label{U2}
\\
K_{t}=U_{t}+2r U_{t}U_{r},\label{Kr2}
\\
K_{r}=U_{r}+r(U_{t}^2+U_{r}^2)+pr e^{2K},\label{Kt2}
\\
K_{rr}-K_{tt}+\frac{U_{r}}{r}+U_{r}^2-U_{t}^2=pe^{2K}, \label{nu2}
\\
K_{r}+\frac{p_{r}}{2p}=0,\label{p2}
\\
K_{t}+\frac{p_{t}}{2p}=0.\label{mu2}
\end{eqnarray}

The energy-momentum conservation equations can be integrated and we 
are left with a 2D wave equation and a quadrature for $K$. 

\begin{eqnarray}
    p=\alpha e^{-2K},\\
K_{t}=U_{t}+2r U_{t}U_{r},\label{Kr}
\\
K_{r}=U_{r}+r(U_{t}^2+U_{r}^2)+\alpha r,\label{Kt}\\
U_{tt}-U_{rr}-\frac{U_{r}}{r}=0.\label{U2D}
\end{eqnarray}

Notice that nice combinations of $K_{t}$ and $K_{r}$ yield squares of 
the derivatives of $U$.

The solution to the 2D wave equation initial data problem provides a 
compact form for writing the general solution,

\begin{equation}
g(r)=U(r,0),\qquad f(r)=U_{t}(r,0),    
    \end{equation}
\begin{equation}
U(r,t)=\frac{1}{2\pi}\int_{0}^{2\pi}d\phi\int_0^1d\tau 
 \frac{\tau}{\sqrt{1-\tau^2}}\left\{tg(v)+f(v)+tf'(v) 
 \frac{t\tau^2+r\tau\cos\phi }{v}
\right\},\end{equation}
where $v=\sqrt{r^2+t^2\tau^2+2rt\tau\cos\phi}$, choosing the origin 
of the polar angle at the angle for $(x,y)$.

There is no need to impose cylindrical symmetry, 
since every solution has an axis provided we define accordingly the 
angle coordinate.

\section{Geodesic completeness}

Provided we use regular initial data, the solution has regular 
components, but this does not guarantee geodesic completeness, as we 
know.

We make use of a theorem devised by us that relates geodesic 
completeness to bounds on the derivatives and the metric functions 
\cite{manolo}:

\begin{description}
\item[Theorem:] A cylindrically symmetric diagonal metric in the form (\ref{metric}) with $C^2$
metric functions  $f,g,\rho$ is future causally geodesically complete 
provided that along causal geodesics:
\begin{enumerate}
\item For large values of $t$ and increasing $r$, 
\begin{enumerate}
    \item $K_{r}+K_{t}\ge 0$, and either $K_{r}\ge 0$ or 
    $|K_{r}|\lesssim K_{r}+K_{t}$.
    \item $(K+U)_{r}+(K+U)_{t}\ge 0$, and either $(K+U)_{r}\ge 0$ 
    or $|(K+U)_{r}|\lesssim (K+U)_{r}+(K+U)_{t}$.
    \item $(K-U-\ln\rho)_{r}+(K-U-\ln\rho)_{t}\ge 0$, and either 
    $(K-U-\ln\rho)_{r}\ge 0$  or $|(K-U-\ln\rho)_{r}|\lesssim 
    (K-U-\ln\rho)_{r}+(K-U-\ln\rho)_{t}$.
\end{enumerate}

\item \label{tt} For  $t$, constant $b$ exists,  
$\left.\begin{array}{c}K(t,r)-U(t,r)\\2\,K(t,r)\\K(t,r)+U(t,r)+\ln\rho(t,r)
\end{array}\right\}\ge-\ln|t|+b.$

\end{enumerate}\end{description}

And similar conditions for past causally geodesically complete 
spacetimes.

These conditions look formidable, but the first set is always 
satisfied for a stiff perfect fluid model \cite{stiff}.

The second set can be seen to reduce to one single condition on the 
axis, namely,

\begin{equation}
U(0,t)=\int_0^1d\tau 
 \frac{\tau}{\sqrt{1-\tau^2}}\left\{tg(|t|\tau)+f(|t|\tau)+|t|\tau 
 f'(|t|\tau) 
\right\}\ge -\frac{1}{2}\ln |t|+b.\label{req}\end{equation} 

\section{Explicit models}

It is not difficult to find examples of functions that fulfill the 
aforementioned condition. Consider for instance polynomials. At $r=0$ 
the equation for $U$ can be integrated. For $f(r)=r^n$, $g(r)=r^m$ we obtain,
\begin{eqnarray}
    U_{f}(t)&=&
    \frac{n!!}{(n-1)!!}\left(\frac{\pi}{2}\right)^{(1+(-1)^{n+1})/2}|t|^n,\\    
    U_{g}(t)&=&\frac{m!!}{(m+1)!!}\left(\frac{\pi}{2}\right)^{(1+(-1)^{m+1})/2}|t|^mt,
\end{eqnarray}
after separating the terms in $f$ and $g$. There are two 
possibilities for a regular model:

\begin{enumerate}
    \item  If $f,g$ are polynomials in $r$ respectively of degree 
    $n,m$ and $n>m+1$, we have a non-singular model if $a_{n}$  is positive.

    \item  If $f,g$ are polynomials in $r$ respectively of degree 
    $n,n-1$, $U_{f}$ and $U_{g}$ at the axis are polynomials of 
    degree $n$ and we have a non-singular model if 
  \begin{equation}
 \left(n+\frac{1}{2}\right)\,a_{n}>|b_{n-1}|. \label{term}\end{equation} 

\end{enumerate}

The set of regular models is certainly not negligible. If we restrict 
ourselves to the space of functions which are polynomials in $t$ at 
$r=0$, we find that the subset of complete models encloses an open 
set, according to the last equation.

\section{Final remarks}

In this talk a large set of cylindrical stiff perfect fluid 
cosmological models has been rederived. An easy sufficient condition 
has been implemented to check whether these models are complete or not. 
Surprisingly the amount of regular models is far larger than expected. 
These results need be extended to more general models, but they may 
point out that complete cosmological models cannot be neglected as 
isolated points in a set of solutions of the Einstein equations.

Pressure seems determinant to avoid singularities in these models. 
The stiff fluid case is a extreme situation where the velocity of 
sound equals the speed of light. On the other hand, pressureless 
models, dust, are known to be singular due to Raychaudhuri equation. 
Barotropic fluids where $p=\gamma \mu$ and $\gamma\in (0,1)$ need be 
explored in more detail.

Nonseparability of the metric functions seems determinant to avoid 
singularities. Separable stiff fluid models had been studied before \cite{agnew} 
and all of them were found to be singular. The non-separable model in 
\cite{leo} already pointed in that direction.



\begin{thebibliography}{99}

\setlength{\itemsep}{-0.6 ex}	
                         
                          
\bibitem{HE} S.W.\ Hawking \& G.F.R.\ Ellis, \textsl{The Large Scale 
Structure of Space-time}, Cambridge University Press 1973

\bibitem{Beem} J.\ Beem, P.\ Ehrlich \& K.\ Easley, \textsl{Global Lorentzian
Geometry\/}, Dekker 1996

\bibitem{Seno} J.M.M.\ Senovilla, \textit{Phys. Rev. Lett.} \textbf{ 
64}, 2219, (1990)\\
F.J. Chinea, L. Fern\'andez-Jambrina, J.M.M. 
Senovilla, \textit{Phys. Rev. D} \textbf{45}, 481 (1992)

\bibitem{esc}  J.M.M.\ Senovilla in: \textit{El Escorial Summer School 
on  Gravitation and General
Relativity 1992: Rotating Objects and Other Topics\/}, Springer-Verlag 1993

\bibitem{manolo}L.\ Fern\'andez-Jambrina \& L.M.\ Gonz\'alez-Romero, 
\textit{Class. Quantum Grav.} \textbf{16}, 953, (1999) [arxiv: 
gr-qc/9812039]\\
L. Fern\'andez-Jambrina,  \textit{Journ. Math. Phys.} 
\textbf{40}, 4028 (1999) [arxiv: gr-qc/9906030]

\bibitem{stiff} L. Fern‡ndez-Jambrina \& L.M. Gonz‡lez-Romero, 
\textit{Phys. Rev. D} \textbf{66}, 024027 (2002) [arxiv: 
gr-qc/0402119]

\bibitem{agnew} A.F. Agnew \& S.W. Goode, \textit{Class. Quantum Grav.} 
\textbf{11}, 1725 (1994)

\bibitem{leo}
L. Fern\'andez-Jambrina, \textit{ Class. Quantum Grav.}, \textbf{ 14}, 3407, 
(1997) [arxiv: gr-qc/0404017].


\end{thebibliography}
\end{document}